\documentclass{article}

\usepackage{microtype}
\usepackage{graphicx}
\usepackage{subfigure}
\usepackage{booktabs}
\usepackage{hyperref}

\usepackage[accepted]{icml2020}

\usepackage{color}
\definecolor{orange}{rgb}{1,0.5,0}
\definecolor{mydarkcyan}{rgb}{0,0.5,0.5}

\icmltitlerunning{Interpreting Stellar Spectra with Unsupervised Domain Adaptation}

\begin{document}

\twocolumn[
\icmltitle{Interpreting Stellar Spectra with Unsupervised Domain Adaptation}

\icmlsetsymbol{equal}{*}
\begin{icmlauthorlist}
\icmlauthor{Teaghan O'Briain}{uvic}
\icmlauthor{Yuan-Sen Ting}{ias,princeton,carnegie,anu}
\icmlauthor{S\'ebastien Fabbro}{uvic,nrc}
\icmlauthor{Kwang Moo Yi}{uvic}
\icmlauthor{Kim Venn}{uvic}
\icmlauthor{Spencer Bialek}{uvic}
\end{icmlauthorlist}

\icmlaffiliation{uvic}{University of Victoria}
\icmlaffiliation{ias}{Institute for Advanced Study, Princeton}
\icmlaffiliation{princeton}{Princeton University}
\icmlaffiliation{carnegie}{Carnegie Observatories}
\icmlaffiliation{anu}{The Australian National University}
\icmlaffiliation{nrc}{NRC Herzberg}

\icmlcorrespondingauthor{S\'ebastien Fabbro}{sfabbro@uvic.ca}

\icmlkeywords{Machine Learning, ICML}

\vskip 0.3in
]

\printAffiliationsAndNotice{} 

\begin{abstract}
We discuss how to achieve mapping from large sets of imperfect simulations and observational data with unsupervised domain adaptation. Under the hypothesis that simulated and observed data distributions share a common underlying representation, we show how it is possible to transfer between simulated and observed domains. Driven by an application to interpret stellar spectroscopic sky surveys, we construct the domain transfer pipeline from two adversarial autoencoders on each domains with a disentangling latent space, and a cycle-consistency constraint. We then construct a differentiable pipeline from physical stellar parameters to realistic observed spectra, aided by a supplementary generative surrogate physics emulator network. We further exemplify the potential of the method on the reconstructed spectra quality and to discover new spectral features associated to elemental abundances.
\end{abstract}

\section{Introduction}
\label{sec:intro}

In an effort to understand the nearby galactic dynamics, and how stellar populations form and evolve, large dedicated sky surveys \citep{buder2018, holtzman2018} are currently accumulating millions of stellar spectra. Large statistical samples of survey data enable galactic archaeologists to construct tools to study the history of galaxy formation or stellar chemical evolution of the Milky Way and the nearby Universe. From the large collections of spectra, estimates of stellar properties such as temperature, surface gravity, metallicities or elemental abundances are routinely produced from data reduction pipelines. The production of these stellar parameter databases requires computationally demanding simulations \citep[{\it e.g.},][]{kurucz1970, meszaros2012new}, and implementing boutique data analysis pipelines \citep[{\it e.g.},][]{perez2016aspcap, ting2019}. However simulations of synthetic spectra are naturally limited to the assumptions behind the stellar physics that we can currently model \citep[{\it e.g.},][]{bialek2020,shetrone2015}. Imperfect corrections of the instrumental and earth-atmosphere signatures also are often left-over systematics from the data-reduction pipelines. It is thus difficult to have a clear understanding of the systematic uncertainties involved with either the data reduction, the synthetic spectral modelling, or from the algorithms limitations.

Thus, in this paper, we propose an unsupervised domain adaptation method, that learns these corrections without human intervention.

\section{Methods}
\label{sec:methods}

Our work is largely based on the \texttt{UNIT} framework \citep{liu2017}, briefly summarized here. The key \texttt{UNIT} idea is to share a latent representation between two unpaired domains of interest. In each domain, a Variational AutoEncoder (VAE) is trained to generate fake samples, sharing the common latent space. The generated samples are thereafter challenged with adversarial classifier networks, trained to discriminate between pairs of reconstructed samples originating from either domain. 

\begin{figure}[ht]
\vskip 0.2in
\begin{center}
\centerline{\includegraphics[width=0.9\columnwidth]{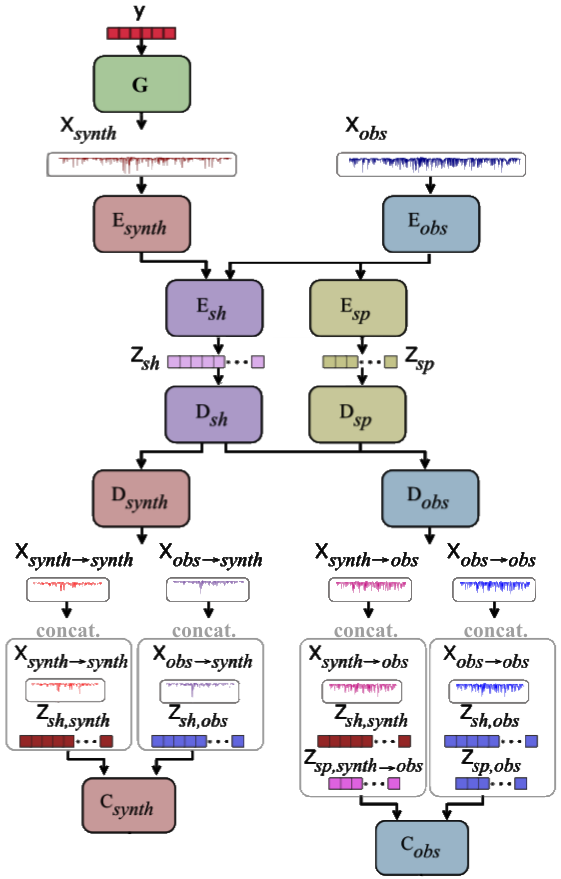}}
\caption{The architecture of the unsupervised domain transfer between simulations and observations. Synthetic spectra are generated on the fly with a physics simulator with input parameters $\theta$. Both synthetic and observed spectra are encoded into a common latent space $\mathcal{Z}_{shared}$. The observed spectra are also encoded simultaneously to a split latent space $\mathcal{Z}_{split}$, coding other observation-specific factors. The rest of the architecture is similar as the UNIT \citep{liu2017} framework.}
\label{fig1}
\end{center}
\vskip -0.3in
\end{figure}

We set the domains of interest to be a large sample of $\mathcal{X}_{synth}$ simulation data and a large sample $\mathcal{X}_{obs}$ of observed data. For our stellar physics purpose, domain data are synthetic and observed spectra. Applying \texttt{UNIT} requires technical modifications to adapt to noisy 1D data, but we also made important architectural choices. We summarize the architecture in the diagram in Fig. \ref{fig1}.  The main high level differences between the \texttt{UNIT} and our current method are: 1) a disentangled latent representation, and 2) stitching a supplementary generative surrogate physics emulator network to the main \texttt{UNIT} framework. Another minor difference is that we have our autoencoders to be deterministic.

\begin{figure}[ht]
\begin{center}
\centerline{\includegraphics[width=1.1\columnwidth]{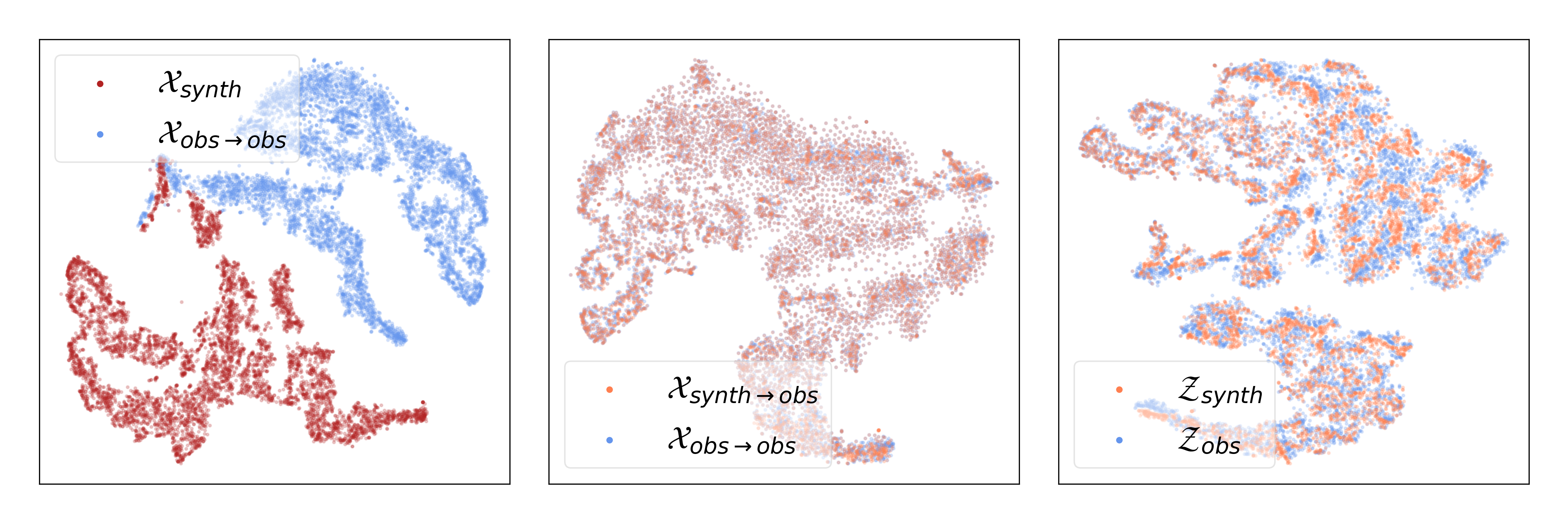}}
\caption{t-SNE where each point represent a spectrum. Left: synthetic and observed spectra. Middle: observed spectra and domain-adapted synthetic spectra by the network. Right: shared latent space.}
\label{fig2}
\end{center}
\vskip -0.3in
\end{figure}

\paragraph{Disentangled Latent Representation}
Even good physics simulators are capable of only partly representing the complexity of all the observations. Thus, it will always suffer from the gap between theory and practice. We model this gap by splitting the latent space in two: one shared latent space between simulations and observations, and one latent space unique to observations. In Fig.~\ref{fig1}, we denote this shared latent space as $\mathcal{Z}_{shared}$ and the latent space unique to observation as $\mathcal{Z}_{split}$. In our experiments, we found that this simple observation was critical to ensure that the shared latent space variable covers the same distributions between the two domains, and contains only the information from stellar physics. To further facilitate sharing of latent space, unlike \texttt{UNIT}, we feed not only the domain translated samples and the decoded spectra, but also their corresponding latent spaces to the discriminators. If the domain transfer succeeds, we expect the shared latent representation to show well-mixed samples originating from both domains, coding physics explainable with the simulator. For our stellar spectra application, we show a t-SNE plot in Fig~\ref{fig2}, where each point represents a spectrum in the respective original domains, in the shared latent space, and after domain adaptation. The t-SNE result shows that the domain-adapted synthetic spectra distribution are well mixed with the observations, closing the synthetic-observation gap.

\paragraph{Surrogate Physics Simulator}
The samples in the synthetic domain are generated by a physical simulator. 
In our application, the physical simulator is computationally expensive, thus producing a stochastic simulator would be very costly. We replace it with a fast neural-network trained to emulate the physics. We carefully selected physical ranges and sampling strategies of the physical parameter space, within the same "expected" range of parameters covered by the observed samples. We then ran the costly simulations, and trained a network capable of emulating the generated samples from the physical parameters. In our particular case, a simple MLP was enough to ensure 0.1\% accuracy at emulating the physical simulator over the full spectral range. Once the differentiable emulator network is trained, the main network can be trained from generating samples on-the-fly.

Besides on-the-fly generation of synthetic spectra, the introduction of the surrogate simulator network allows us to constitute an end-to-end system from interpretable physical parameters to realistic spectra which implicitly learns the non-modelled physics. Being end-to-end, the system is also fully differentiable, and allows one to differentiate the the domain-adapted synthetic spectra with respect to physical parameters. We, in fact, utilize this property to infer physics from spectra.

\section{Experiments}
\label{sec:experiments}
We present two case studies to demonstrate how we generate systematic-corrected synthetic models from unlabelled observed spectra through domain adaptation.

\subsection{Data}
For these two experiments, we use the publicly available APOGEE \cite{holtzman2018} survey data, which comprises of 250,000 high-resolution $R \equiv \lambda/\Delta \lambda \sim 22,000$ infrared spectra, where $\lambda$ designates wavelength.

\begin{itemize}
\item For the synthetic domain, we use 
the Kurucz {\sc atlas12}/{\sc synthe} models \citep[see][for details]{ting2019} 
\item As for the observed domain, we adopt the APOGEE Data Release 14 spectra. The APOGEE spectra are wavelength calibrated to vacuum to be consistent with the Kurucz models. Multiple visits to the same object are pre-processed and co-added. We corrected the spectra for spectral redshift and self-consistently continuum normalized both the Kurucz models and APOGEE spectra using the same routine as laid out in \citet{ting2019}. 
\end{itemize}

The data set provides estimated stellar parameters for both dwarf and giant stars. We further use the samples in the dataset in a random order to make sure spectra from the two domains are not paired.

APOGEE spectra are adopted from half of the objects as the observed domain, and for the other half, we generate synthetic Kurucz spectra via the surrogate emulator assuming only their stellar parameters. We inherit the 25 stellar parameters derived in \citet{ting2019}: $T_{\rm eff}$, $\log g$, microturbulence $v_{\rm turb}$, additional broadening $v_{\rm broad}$, and 20 elemental abundances, namely, C, N, O, Na, Mg, Al, Si, P, S, K, Ca, Ti, V, Cr, Mn, Fe, Co, Ni, Cu, Ge, and the isotopic ratio, C12/C13. This procedure yields a training set of about 100,000 spectra in each domain. We adopt 80,000 spectra as the training set, 10,000 spectra as the validation set.  The rest (7,000 spectra) are held out as the test set. All results below are based on the test set.

\subsection{Stellar Spectra Translation}

As a natural application, we investigate the quality of the translated spectra in the observed domain to that originated from the synthetic domain.
Once the framework is trained, the generation is fast and allows us to perform a maximum-likelihood fit between the translated and the observed spectra to get the best-fit stellar parameters. 

\begin{figure}[ht]
\begin{center}
\centerline{\includegraphics[width=1.1\columnwidth]{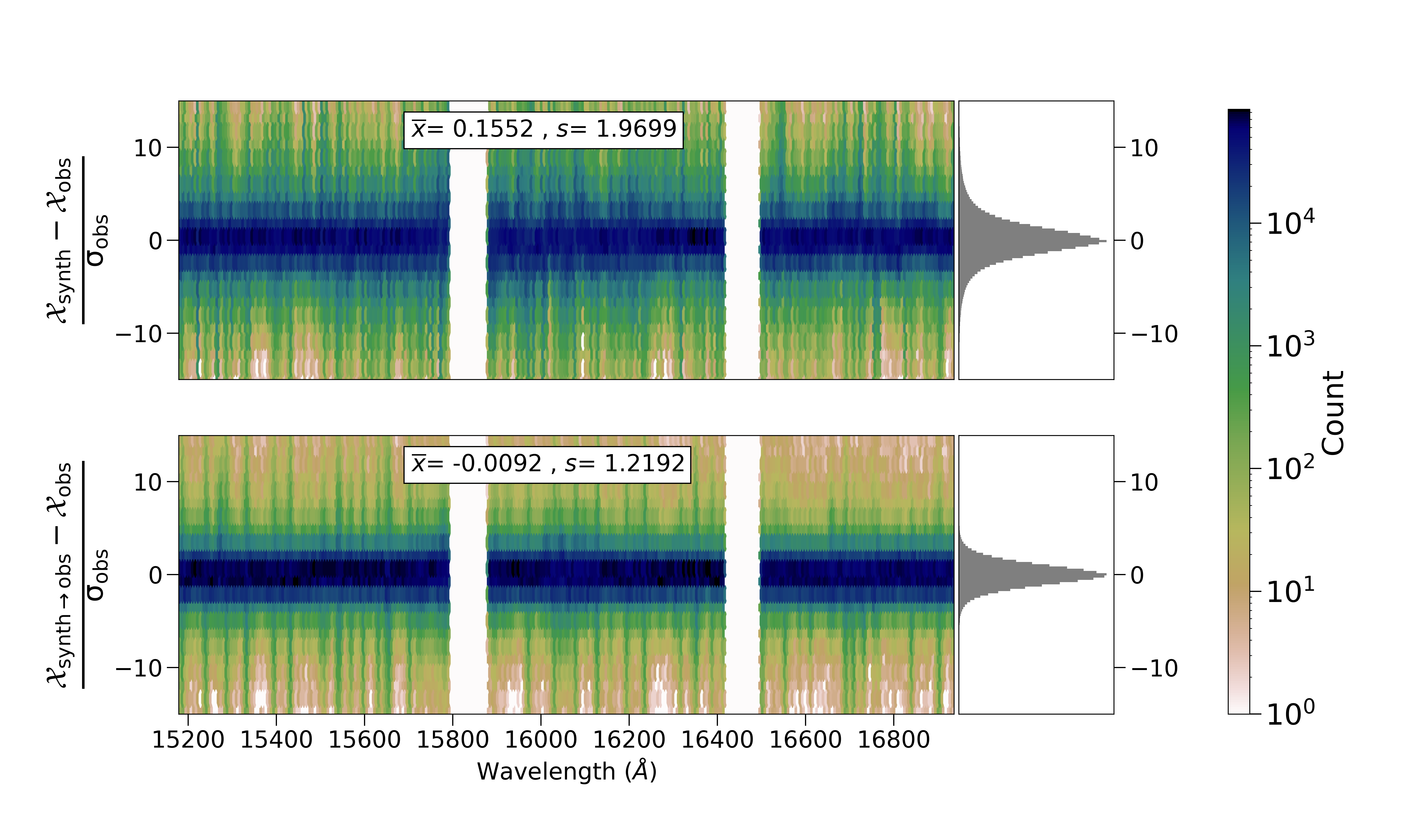}}
\caption{Residuals between physical (Kurucz) best-fit modelled spectra and the APOGEE observed spectra. The top panel shows the difference between the 10,000 APOGEE test spectra and their corresponding best-fit Kurucz models. The bottom panel is a similar comparison but with the transferred spectra by our network, showing smaller mean bias $\overline{x}$ and standard deviation $s$ of the residuals.}
\label{fig3}
\end{center}
\vskip -0.3in
\end{figure}

In Fig~\ref{fig3}, we compare the average residuals between our transferred best-fit model (bottom) to a best-fit emulated (non-translated) physics model (top).
Fitting spectra is a typical process for stellar parameter analysis.
The residuals are normalised with the spectra uncertainties provided by the APOGEE survey data release.
Well calibrated uncertainties with unbiased best-fit models would show a distribution of $z$-score residuals following $\mathcal{N}(0,1)$. We show the sample mean $\overline{m}$ and standard deviation $s$ in each case. The systematics-learned model clearly outperforms the stellar physics-only model.

\subsection{Spectral Lines}
Here, we study the derivatives of the full network, in particular the derivatives with respect to elemental abundances. The derivatives ({\it i.e.}, flux ``response'') of elemental abundances examine which spectral features are associated with a particular element. Unfortunately, for the APOGEE observed spectra, it is impossible to know the ground truth --  we simply do not know what might be missing in the Kurucz models. Therefore, as a proof of concept, a mock ``observed'' data set is drawn from the Kurucz models instead, of which we know the ground truth derivatives. More advanced applications are deferred to future studies.

\begin{figure}[ht]
\begin{center}
\centerline{\includegraphics[width=\columnwidth]{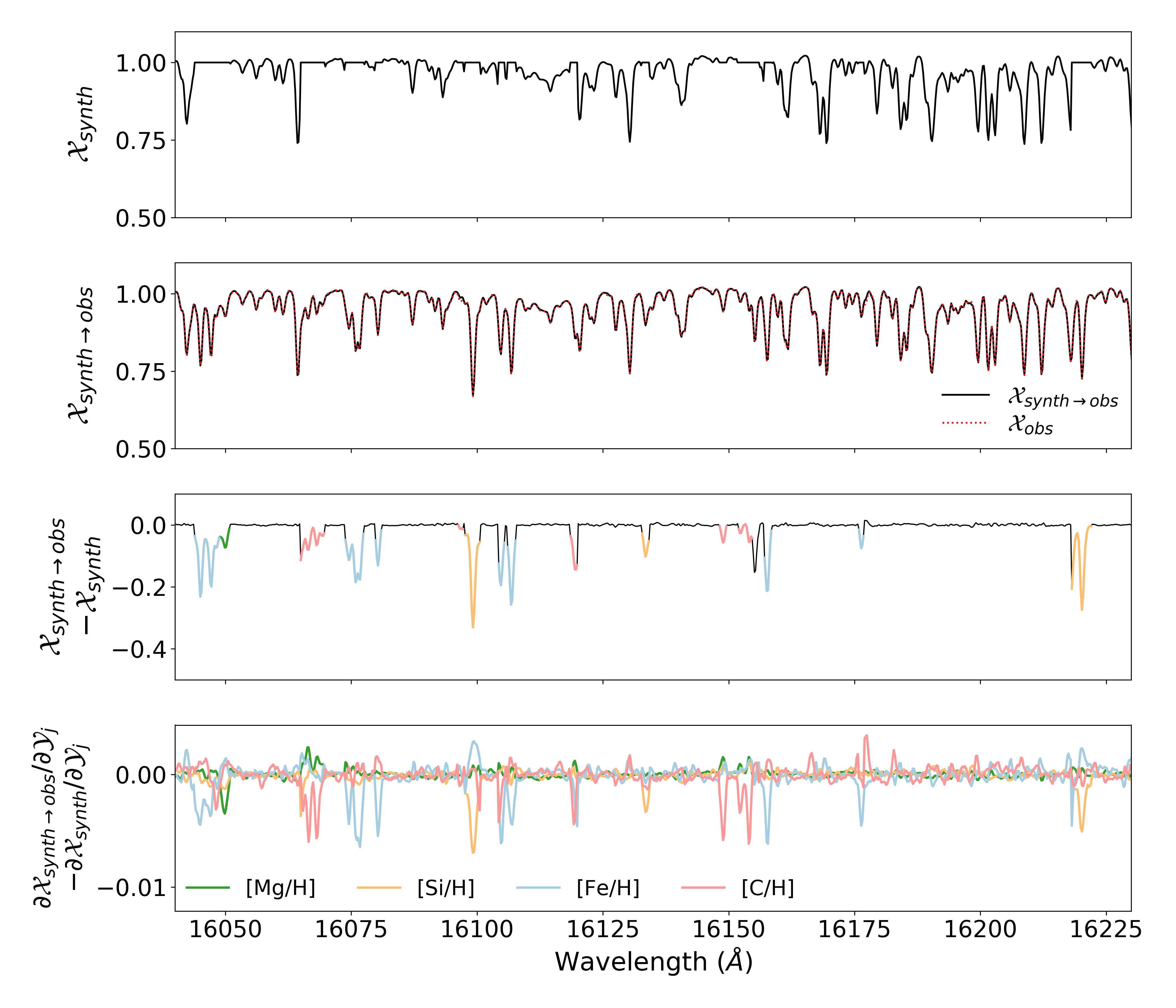}}
\caption{Generated Kurucz spectral models but 30\% masked of the spectral features in the “synthetic” domain, and original Kurucz models (with noise) as the observed domain. The top panel shows the Kurucz spectrum with missing spectral features, whereas the second panel shows the systematic-corrected synthetic spectra. The third panel shows the differences between the synthetic and “observed” spectrum, demonstrating the missing features; the missing features of Mg, Si, Fe, and C are annotated in blue, orange, green, and red respectively. The final panel shows the differences in the derivatives between the synthetic domain and the transferred models.}
\label{fig4}
\end{center}
\vskip -0.3in
\end{figure}

For the observed domain, instead of adopting the APOGEE spectra, an ``observed'' data set is synthesized with our physics simulator. We further add noise to these observed mock spectra to mimic the same noise distribution of the APOGEE spectra. As for the synthetic domain, we mask approximately 30\% of the spectral features randomly by setting them to the continuum level.
In short, in this controlled experiment, we assume two sets of unpaired Kurucz models.

To demonstrate that we can learn actual physics, the goal here is to correctly identify the physical source ({\it i.e.}, which element) of these masked spectral features, which act as missing information in the synthetic models that we want to learn and recover from the data.

We calculate the difference between the flux derivatives of the synthetic emulator, $\partial G(\mathcal{Y})/\partial \mathcal{Y}$ and the derivatives of the domain transferred spectra, $\partial T_{synth\rightarrow obs}(G(\mathcal{Y}))/\partial \mathcal{Y}$, where $G$ is the surrogate emulator, and $T_{synth\rightarrow obs}$ the domain adaptation network and $\mathcal{Y}$ the physical stellar parameters. The former informs us about the original spectral features in the ``synthetic'' domain, which has about 30\% missing spectral features. The latter reveals the ``true'', unmasked, Kurucz line list learned from the ``observed'' data.

Fig.~\ref{fig4} shows the result of domain adaptation. We assume a typical K-giant star with Solar elemental abundances to be the reference point. The top panel shows the spectrum from the synthetic domain, and the second panel shows the transferred spectrum. By construction, the synthetic spectrum is missing many spectral features compared to the transferred spectrum. Although not shown, the transferred spectrum is almost identical to the observed spectrum in this case. When mapped to the observed domain, missing features in the synthetic domain are faithfully reconstructed.

The last two panels demonstrate that not only we can faithfully reconstruct the missing features, but it also correctly links them to their corresponding elements. The third panel shows differences between the synthetic spectrum and the transferred spectrum. The masked features are colour-coded with their associated elements and we focus on the three elements that have a prominent presence in the APOGEE spectra, namely Mg, Si, Fe, and C. The final panel shows the differences between the true and recovered derivatives. As shown in this final panel, the differences in the derivatives are the strongest when calculated with the correct input element. The missing features are recovered with accurate associations, even though the training was trained with noisy, unpaired, observed spectra that mimic the APOGEE observations.

\section{Conclusion}
\label{sec:discussion}

In this study, we illustrated that domain-adaptation framework can have broad implications for simulation-based physics. As a proof of concept, we focused on stellar spectral modeling. Synthetic spectral models are auto-calibrated through a domain adaptation network, thereby reducing the gap between theory and observations. The same network can also predict the elemental sources of unknown spectral features in the synthetic models.

\bibliography{refs}
\bibliographystyle{icml2020}

\end{document}